

\magnification=\magstep1
\baselineskip=18pt
\nopagenumbers
\line{\hfil CU-TP-688}
\line{\hfil gr-qc/9504003}
\vglue .5in
\centerline{\bf Are All Static Black Hole Solutions Spherically Symmetric?}
\vskip .3in
\centerline{\it S.~Alexander Ridgway and Erick J.~Weinberg}
\vskip .1in
\centerline{Physics Department, Columbia University}
\centerline{New York, New York 10027}
\vskip .4in
\baselineskip=14pt
\overfullrule=0pt
\centerline {\bf Abstract}

The static black hole solutions to the Einstein-Maxwell equations are
all spherically symmetric, as are many of the recently discovered
black hole solutions in theories of gravity coupled to other forms of
matter.  However, counterexamples demonstrating that static black
holes need not be spherically symmetric exist in theories, such as the
standard electroweak model, with electrically charged massive
vector fields.  In such theories, a magnetically charged
Reissner-Nordstr\"om solution with sufficiently small horizon radius
is unstable against the development of a nonzero vector field
outside the horizon.  General arguments show that, for generic values
of the magnetic charge, this field cannot be spherically symmetric.
Explicit construction of the solution shows that it in fact has no
rotational symmetry at all.

\vskip .1in
\noindent\footnote{}{\noindent This work was supported in part by
the US Department of Energy.}

\vfill\eject
\baselineskip=18pt

\def\pr#1#2#3#4{{\it Phys. Rev. D\/}{\bf #1}, #2 (19#3#4)}
\def\cmp#1#2#3#4{{\it Commun. Math. Phys.} {\bf #1}, #2 (19#3#4)}
\def\np#1#2#3#4{{\it Nucl. Phys.} {\bf B#1}, #2 (19#3#4)}

\def\prl#1#2#3#4{{\it Phys. Rev. Lett.} {\bf #1}, #2 (19#3#4)}

      One of the many remarkable aspects of black holes is the high
degree of symmetry of the classical black hole solutions.  Both of the
static solutions that were discovered in the early days of general
relativity --- the Schwarzschild and the Reissner-Nordstr\"om --- are
spherically symmetric.  At one time, one might have thought that this
simply reflected the fact that spherically symmetric solutions are
easier to find.  However, it was shown two decades ago [1]
that these are in fact the only static electrovac black hole
solutions.  Does this result generalize to gravity
coupled to other types of matter; i.e., are static black
holes always spherically symmetric?  The answer [2], as we will describe
below, is no.

     The restrictions on the possible electrovac black holes can be
viewed as just one instance of the ``no-hair'' results that limit the
possible structure of black holes in a number of matter theories.
This suggests that in seeking solutions that depart from spherical
symmetry one should look to theories that do not admit no-hair
theorems; these tend to be [3] theories that possess static soliton
solutions in the absence of gravity.
An important class of such theories is the spontaneously broken
gauge theories that possess nonsingular magnetic monopole solutions
[4] in
the absence of gravity; the simplest example is the SU(2) gauge
theory with the symmetry broken to the U(1) of electromagnetism by a
triplet Higgs field.  The elementary particles of this theory include,
in addition to the massless photon, a spin-one particle with mass $m$
and electric charge $e$ and an electrically neutral spinless particle.

This theory possesses spherically symmetric black hole solutions [5] with
nontrivial matter fields outside the horizon; these can be constructed
by numerical
integration of the field equations.  They carry magnetic
charge $1/e$, and may be viewed as Schwarzschild-like black holes
embedded in the center of~'t~Hooft-Polyakov monopoles.  In fact, one
does not need a detailed examination of the field equations to show
that such solutions exist.  An analysis [6] of the small fluctuations
about the Reissner-Nordstr\"om solution reveals an instability leading
to the development of massive vector meson fields just outside the
horizon whenever the horizon radius is less than a critical value of
order $m^{-1}$; the existence of this instability indicates that there
must be a nearby static solution with hair.

      The physical basis for this instability is easily understood.
Charged particles with nonzero spin in general have magnetic
moments.  In a sufficiently strong magnetic field, the energetic cost
of producing a cloud of such particles can be more than offset by
the energy gained by aligning their magnetic moments so as to
partially shield the magnetic field.

Hence, the essential features needed for these new black holes are
captured by a theory that includes, in addition to the electromagnetic
and gravitational fields, a massive charged vector field $W_\mu$ with
a magnetic moment fixed by an arbitrary parameter $g$.
This theory has the flat spacetime Lagrangian
$$  \eqalign{ {\cal L} &= -{1\over 4}F_{\mu\nu}F^{\mu\nu}
        -{1\over 2} | D_\mu W_\nu -D_\nu W_\mu |^2
        -  m^2 W^*_\mu W^\mu \cr   &\qquad
     -{ieg\over 4} F^{\mu\nu} \left( W^*_\mu W_\nu -W^*_\nu W_\mu\right)
     -{\lambda e^2 \over 4} \left| W^*_\mu W_\nu -W^*_\nu W_\mu\right|^2}
   \eqno(1) $$
where $F_{\mu\nu}$ is the electromagnetic field strength and $D_\mu=
\partial_\mu -ieA_\mu$ denotes the electromagnetic gauge covariant derivative.
For $g=2$ and $\lambda=1$ this theory is essentially the spontaneously
broken gauge theory discussed above, but with the terms involving the
scalar field omitted.  By setting $g=2$ and $\lambda = 1/\sin^2
\theta_W$, we obtain instead a portion of the standard electroweak
theory.

The electromagnetic vector potential must have a singularity on any
surface enclosing a magnetic charge.  In order that the resulting
``Dirac string'' singularity not be physically detectable by charged
particles, the magnetic charge must be equal to $q/e$, with $q$ either an
integer or a half-integer.  The black holes with $q=1$ are just the
spherically symmetric solutions described above.

      Let us now consider magnetically charged black holes with $q\ne
1$ in the context of this theory.  For any allowed value of $q$ there
is a Reissner-Nordstr\"om solution for which the $W$ field vanishes
identically.  Analysis [7] of the small fluctuations about this
solution shows an instability similar to that found in the
$q=1$ case.  Hence, we are once again led to the
existence of new black hole solutions with hair.  What is new here is
that these solutions cannot be spherically symmetric.  To understand
this, note that any static field can be expanded as
a sum of generalized spherical harmonics multiplied by
functions of $r$.  These
spherical harmonics are the eigenfunctions of the quantum
mechanical angular momentum operator; a configuration is spherically
symmetric if its expansion contains only harmonics with zero angular
momentum.
The harmonics used in the expansion must be must be appropriate
to the type of field.  For an electrically charged vector field in the
presence of a magnetic monopole with magnetic charge $q/e$, one needs
monopole vector harmonics [8].  These are the eigenfunctions of the total
angular momentum operator for a spin-one particle with electrical
charge $e$ moving in the presence of the monopole.  In addition to the
usual orbital angular momentum ${\bf r} \times m{\bf v}$ and the spin
angular momentum, there is an anomalous angular momentum of magnitude
$q \hbar$ oriented along the line from the electric charge to the
monopole.  Because this anomalous contribution is perpendicular to the
orbital angular momentum, it is impossible to have vanishing total
angular momentum if the spin angular momentum is smaller than the
anomalous component.  Hence, if $q > 1$ there is no monopole vector
harmonic with zero angular momentum.

     Because these new solutions are not spherically symmetric, one
would expect them to be quite difficult to obtain.  However, a
perturbative approach is available for the case where the
Reissner-Nordstr\"om solution is just barely unstable.  In this
situation, one would expect the unstable modes to give an indication
of the nature of the nearby stable solution.  This suggests that we
begin by identifying the terms in the energy for static configurations
that are quadratic in the perturbations about the Reissner-Nordstr\"om
solution.  This gives a quadratic form that has a single negative
eigenvalue $-\beta^2 m^2$, where $\beta$ tends to zero as the horizon
radius approaches the critical value for instability.  Because the
corresponding modes, which involve only $W_\mu$, cannot be spherically
symmetric, they must form a degenerate multiplet corresponding to an
irreducible representation of the rotation group; let us label these
$\psi_\mu^M$.

    We now write the vector field as a linear combination of these
functions plus a remainder that is of higher order in $\beta$:
$$  W_\mu = \sum_M k_M \psi_\mu^M + \tilde W_\mu \, .
    \eqno(2) $$
Next, the field equations can be used to determine the leading order
deviations of the electromagnetic field and of the metric from their
Reissner-Nordstr\"om values in terms of $W_\mu$.  The resulting
expressions, together with Eq.~(2), must then be substituted back into
the energy.  The dominant terms in the energy then become a fourth
order polynomial in the $k_M$, whose minimum determines the values of
the $k_M$, and thus the leading approximation to the solution.

    This program is particularly simple to carry out if $g>0$ and $q
\ge 1$.  In this case, the $\psi_\mu^M$ form a multiplet of $2q-1$
functions of the form
$$  \psi_t =0, \qquad \qquad
 \psi_j^M  = f(r) C_j^{q-1, M}(\theta,\phi)
    \eqno(3) $$
where the $C_j^{q-1, M}(\theta,\phi)$ are vector spherical harmonics
with total angular momentum quantum number $J=q-1$ and $f(r)$ is a
function that is nonzero on the horizon and falls exponentially with
distance for $r > m^{-1}$.  Hence, the vector field outside the
horizon is of the form
$$ W_\mu = f(r) \Phi_\mu(\theta,\phi) \, .
    \eqno(4)$$
Using the explicit forms for the monopole harmonics, one can show that
$\Phi^*_\mu\Phi^\mu$ has $2(q-1)$ zeros on the unit sphere.
This fact by
itself makes it quite clear that spherical symmetry is impossible for
$q \ne 1$.  One could conceivably have axial symmetry, with the zeros
either all coinciding or else lying at two antipodal points.  However,
detailed study of the solutions shows that this is not what happens.
Instead, the zeros tend to be distributed as evenly as possible over
the unit two-sphere, leading to a configuration without any rotational
symmetry.
This vector field configuration induces higher multipole
components in the electromagnetic field and the metric, with
multipoles of order up to $2(q-1)$.  At large distances from the black
hole these fall as appropriate powers of $r$.

    At one time the known static black hole solutions were all algebraically
simple and highly symmetric.  In recent years we have learned that a
number of types of matter allow black holes that are considerably more
complex in that they have nontrivial fields outside the horizon.  The
work described here shows that the inclusion of magnetic charge
gives rise to objects with even more structure --- black
holes that have no rotational symmetry at all.

\bigskip
\centerline{\bf References}
\medskip
\noindent 1.  W.~Israel, \cmp 8{245}68.

\noindent 2.  S.A.~Ridgway and E.J.~Weinberg, Columbia preprint CU-TP-673,
gr-qc/9503035.

\noindent 3.  D.~Kastor and J.~Traschen, \pr {46}{5399}92.

\noindent 4.  G.~'t~Hooft, \np {79}{276}74; A.M.~Polyakov, {\it Pisma
v. Zh. E.T.F.,} {\bf 20}, 430 (1974) [{\it JETP Lett.} {\bf 20}, 194
(1974)].

\noindent 5.  K.~Lee, V.P.~Nair and E.J.~Weinberg, \pr {45}{2751}92;
P.~Breitenlohner, P.~Forg\'acs, and D.~Maison, \np {383}{357}92.

\noindent 6.  K.~Lee, V.P.~Nair and E.J.~Weinberg, \prl {68}{1100}92.

\noindent 7.  S.A.~Ridgway and E.J.~Weinberg, \pr {51}{638}95.

\noindent 8.  H.A.~Olsen, P.~Osland, and T.T.~Wu, \pr {42}{665}90;
E.J.~Weinberg, \pr {49}{1086}94.

\end